\documentclass[prc,twocolumn,superscriptaddress,twoside, %
floatfix]{revtex4}

\usepackage{amssymb,epsfig}
\usepackage{xcolor}

\begin{document}

\title{Observation of the $^7$H excited state}

\author{A.A.~Bezbakh}
\affiliation{Flerov Laboratory of Nuclear Reactions, JINR,  141980 Dubna,
Russia}
\affiliation{Institute of Physics, Silesian University in Opava, 74601  Opava, Czech Republic}

\author{V.~Chudoba}
\email{chudoba@jinr.ru}
\affiliation{Flerov Laboratory of Nuclear Reactions, JINR,  141980 Dubna,
Russia}
\affiliation{Institute of Physics, Silesian University in Opava, 74601  Opava, Czech Republic}

\author{S.A.~Krupko}
\affiliation{Flerov Laboratory of Nuclear Reactions, JINR,  141980 Dubna,
Russia}
\affiliation{SSC RF ITEP of NRC ``Kurchatov Institute'', 117218 Moscow, Russia}

\author{S.G.~Belogurov}
\affiliation{Flerov Laboratory of Nuclear Reactions, JINR,  141980 Dubna,
Russia}
\affiliation{National Research Nuclear University ``MEPhI'',
	115409 Moscow, Russia}

\author{D.~Biare}
\affiliation{Flerov Laboratory of Nuclear Reactions, JINR,  141980 Dubna,
Russia}

\author{A.S.~Fomichev}
\affiliation{Flerov Laboratory of Nuclear Reactions, JINR,  141980 Dubna,
Russia}
\affiliation{Dubna State University, 141982 Dubna, Russia}

\author{E.M.~Gazeeva}
\affiliation{Flerov Laboratory of Nuclear Reactions, JINR,  141980 Dubna,
Russia}


\author{A.V.~Gorshkov}
\affiliation{Flerov Laboratory of Nuclear Reactions, JINR,  141980 Dubna,
	Russia}

\author{L.V.~Grigorenko}
\affiliation{Flerov Laboratory of Nuclear Reactions, JINR,  141980 Dubna,
Russia}
\affiliation{National Research Nuclear University ``MEPhI'',
115409 Moscow, Russia}
\affiliation{National Research Centre ``Kurchatov Institute'', Kurchatov
sq.\ 1, 123182 Moscow, Russia}

\author{G.~Kaminski}
\affiliation{Flerov Laboratory of Nuclear Reactions, JINR,  141980 Dubna,
Russia}
\affiliation{Heavy Ion Laboratory, University of Warsaw, 02-093 Warsaw, Poland}

\author{O.~Kiselev}
\affiliation{GSI Helmholtzzentrum f\"ur Schwerionenforschung GmbH, 64291 Darmstadt, Germany}

\author{D.A.~Kostyleva}
\affiliation{GSI Helmholtzzentrum f\"ur Schwerionenforschung GmbH, 64291 Darmstadt, Germany}
\affiliation{II. Physikalisches Institut, Justus-Liebig-Universit\"at, 35392
Giessen, Germany}

\author{M.Yu.~Kozlov}
\affiliation{Laboratory of Information Technologies, JINR,  141980 Dubna,
	Russia}

\author{B. Mauyey}
\affiliation{Flerov Laboratory of Nuclear Reactions, JINR,  141980 Dubna,
	Russia}

\author{I.~Mukha}
\affiliation{GSI Helmholtzzentrum f\"ur Schwerionenforschung GmbH, 64291 Darmstadt, Germany}

\author{I.A.~Muzalevskii}
\affiliation{Flerov Laboratory of Nuclear Reactions, JINR,  141980 Dubna,
Russia}
\affiliation{Institute of Physics, Silesian University in Opava, 74601  Opava, Czech Republic}

\author{E.Yu.~Nikolskii}
\affiliation{National Research Centre ``Kurchatov Institute'', Kurchatov
sq.\ 1, 123182 Moscow, Russia}
\affiliation{Flerov Laboratory of Nuclear Reactions, JINR,  141980 Dubna,
Russia}

\author{Yu.L.~Parfenova}
\affiliation{Flerov Laboratory of Nuclear Reactions, JINR,  141980 Dubna,
Russia}

\author{W.~Piatek}
\affiliation{Flerov Laboratory of Nuclear Reactions, JINR,  141980 Dubna, 
	Russia}
\affiliation{Heavy Ion Laboratory, University of Warsaw, 02-093 Warsaw, Poland}

\author{A.M.~Quynh}
\affiliation{Flerov Laboratory of Nuclear Reactions, JINR,  141980 Dubna,
Russia}
\affiliation{Nuclear Research Institute, 670000 Dalat, Vietnam}

\author{V.N.~Schetinin}
\affiliation{Laboratory of Information Technologies, JINR,  141980 Dubna,
	Russia}

\author{A.~Serikov}
\affiliation{Flerov Laboratory of Nuclear Reactions, JINR,  141980 Dubna,
Russia}

\author{S.I.~Sidorchuk}
\affiliation{Flerov Laboratory of Nuclear Reactions, JINR,  141980 Dubna,
Russia}

\author{P.G.~Sharov}
\affiliation{Flerov Laboratory of Nuclear Reactions, JINR,  141980 Dubna,
Russia}
\affiliation{Institute of Physics, Silesian University in Opava, 74601  Opava, Czech Republic}

\author{R.S.~Slepnev}
\affiliation{Flerov Laboratory of Nuclear Reactions, JINR,  141980 Dubna,
Russia}

\author{S.V.~Stepantsov}
\affiliation{Flerov Laboratory of Nuclear Reactions, JINR,  141980 Dubna,
Russia}

\author{A.~Swiercz}
\affiliation{Flerov Laboratory of Nuclear Reactions, JINR,  141980 Dubna,
Russia}
\affiliation{AGH University of Science and Technology, Faculty of Physics and
Applied Computer Science, 30-059 Krakow, Poland}

\author{P.~Szymkiewicz}
\affiliation{Flerov Laboratory of Nuclear Reactions, JINR,  141980 Dubna,
Russia}
\affiliation{AGH University of Science and Technology, Faculty of Physics and
Applied Computer Science, 30-059 Krakow, Poland}

\author{G.M.~Ter-Akopian}
\affiliation{Flerov Laboratory of Nuclear Reactions, JINR,  141980 Dubna,
Russia}
\affiliation{Dubna State University, 141982 Dubna, Russia}

\author{R.~Wolski}
\affiliation{Flerov Laboratory of Nuclear Reactions, JINR,  141980 Dubna,
Russia}

\author{B.~Zalewski}
\affiliation{Flerov Laboratory of Nuclear Reactions, JINR,  141980 Dubna,
Russia}
\affiliation{Heavy Ion Laboratory, University of Warsaw, 02-093 Warsaw, Poland}

\author{M.V.~Zhukov}
\affiliation{Department of Physics, Chalmers University of Technology, S-41296
G\"oteborg, Sweden}

\date{\today.}

\begin{abstract}
	The $^7$H system was populated in the $^2$H($^8$He,$^3$He)$^7$H reaction with a 26 AMeV $^8$He beam.
	The $^{7}$H missing mass energy spectrum, the $^{3}$H energy and angular distributions in the $^7$H decay frame were reconstructed.
	The $^7$H missing mass spectrum shows a peak which can be interpreted either as unresolved $5/2^+$ and $3/2^+$ doublet or one of these states at 6.5(5) MeV.
	The data also provide indications on the $1/2^+$ ground state of $^7$H located at 2.0(5) MeV with quite a low population cross section of $\sim 10$ $\mu$b/sr within angular range $\theta_{\text{cm}} \simeq 6^{\circ} - 30^{\circ}$.
\end{abstract}


\keywords{\textcolor{red}{XXX, YYY, ZZZ}}

\maketitle

\textit{Introduction.} --- %
%
	The $^{7}$H nucleus is a special system in the ``world of nuclides''.
	It is the heaviest conceivable hydrogen isotope with the largest $A/Z=7$ ratio, which is closer to ``neutron matter'' than any other known nuclide.
	The closed $p_{3/2}$ neutron subshell of its ground state (g.s.) implies special stability relative to its isobaric neighbors.
	The $^{7}$H g.s.\ decays via unique five-body $^{3}$H+$4n$ decay channel.
	This form of nuclear dynamics has not yet been studied at all, and it was discussed that in $^7$H this decay mechnism may lead to such an exclusive phenomenon as $4n$ radioactivity \cite{Golovkov:2004,Grigorenko:2011}.
	Unfortunately, there is no definite reliable information about such an interesting system.
	
	The search for $^{7}$H has a long, but not fortunate history.
	It was searched but not found among the nuclear-stable products of ternary fission of $^{252}$Cf \cite{Aleksandrov:1982} and in pion double charge-exchange $^7$Li($\pi^-$,$\pi^+$) reaction \cite{Gornov:2003}.
	Since the emergence of the radioactive ion beams (RIB), the evident way to search for $^{7}$H is proton removal from $^{8}$He.
	The $^1$H($^{8}$He,$^{2}$He) reaction was used in Ref.\ \cite{Korsheninnikov:2003} and evidence for intense population of $^{7}$H spectrum right above the $^{3}$H+$4n$ threshold was demonstrated.
	Low energy resolution ($1.9$ MeV) and high background did not allow to draw a quantitative conclusion in this work.
	The $^2$H($^{8}$He,$^{3}$He) reaction at 21 AMeV on a thick cryogenic deuterium target was used in Ref.\ \cite{Golovkov:2004} for the specific task of searching for extreme low-lying (and therefore long-living) $^{7}$H g.s.
	Together with theoretical estimates for the lifetimes in the five-body decays, this allowed to establish the lower decay-energy limit $E_T > 50-100$ keV for $^7$H.
	The decay energy $E_T$ and missing mass (MM) mean the same value having zero value at the $^{3}$H+$4n$ decay threshold.
	The observation of a quite low-lying $^{7}$H resonance state with $E_T \sim 0.57$ MeV produced in the $^{12}$C($^{8}$He,$^{7}$H)$^{13}$N reaction was declared in Ref. \cite{Caamano:2007}.
	An important deficiency of this work was the difficulty of the reaction-channel identification.
	The observed events could belong also to $^{6}$H or to $^{5}$H continuum.
	The next attempt to obtain $^7$H was made using the  $^2$H($^{8}$He,$^{3}$He) reaction carried out at the $^8$He projectile energy 42 AMeV \cite{Nikolskii:2010}.
	Quite a smooth excitation spectrum was obtained in this work
	and authors pointed out a peculiarity at $E_T \sim$ 2 MeV at a cross-section level of $\sim 30$ $\mu$b/sr. 

	Though the $^{7}$H production from $^{8}$He seems to be a straightforward idea, it had not provided a decisive result within the last 15 years of research.
	In the present work we for the first time obtain a reliable quantitative results for the $^{7}$H energy spectrum coming closer to the solution of the $^{7}$H g.s.\ problem.

	\begin{figure}[t]
	\begin{center}
	\includegraphics[width=0.49\textwidth]{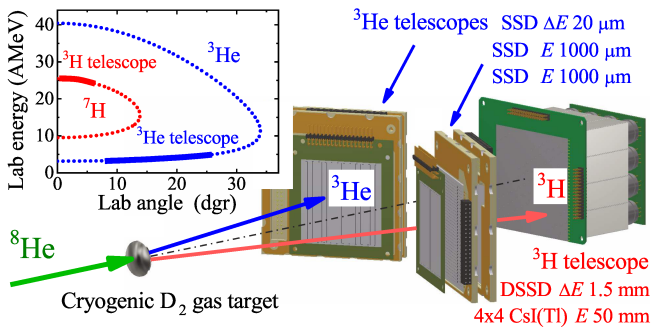}
	\end{center}
	\caption{The sketch of the experimental setup. The inset shows kinematical conditions for the $^2$H($^8$He,$^3$He)$^7$H reaction at 26 AMeV.}
	\label{fig:setup}
	\end{figure}

\textit{Experiment.} --- %
%
	It was performed at the Flerov Laboratory of Nuclear Reactions (JINR) at the ACCULINNA-2 fragment separator \cite{Fomichev:2018}.
	This facility was commissioned in 2017, and this run was the first one performed with the full intensity primary beam.
	The 33.4 AMeV $^{11}$B beam was delivered by the U-400M cyclotron with the intensity of about 1 p$\mu$A. It was focused in the 5-mm spot on the 1 mm thick beryllium production target.
	The secondary $^8$He beam with energy of $26$ AMeV and $\sim 90\%$ purity, having intensity of $\sim 10^5$ pps, was focused into a 17-mm spot on the deuterium gas target. The $D_2$ target was cooled to 27 K, and its thickness  made $\sim 3.8\times10^{20}$ cm$^{-2}$.
	Beam tracking was provided by two multi-wire proportional chambers located by 27 and 82 cm upstream the $D_2$ target and giving the individual $^{8}$He hit positions on the target with 1-mm accuracy.
	The time-of-flight detector system, which identified each particle in the secondary beam and measured its energy, consisted of two thin plastic scintillators with 12.3 m flight path having 0.2 ns time resolution.

	The experimental setup is shown in Fig.\ \ref{fig:setup}.
	Choosing the same ($d$,$^{3}$He) reaction as in \cite{Nikolskii:2010}, we, however, had to optimize the setup in a different way.
	Energy resolution for the $^7$H missing mass measurement, estimated by Monte-Carlo method, at a level of $\sim 1.1$ MeV which is two times better than in \cite{Nikolskii:2010}. 
	A set of the two identical $\Delta E$-$E$-$E$ telescopes was the key installation of the experiment destined to detect the low-energy $^3$He recoil nuclei emitted in the $^2$H($^8$He,$^3$He)$^7$H reaction in the range of $9-20$ MeV.
	Each telescope consisted of three Si strip detectors --- one 20-micron SSD ($50\times 50$ mm, 16 strips) and two 1000-micron SSDs ($61\times 61$ mm$^2$, 16 strips), where the second 1000-micron detector operated as veto.
	The telescopes were located $166$ mm downstream from the $D_2$ target covering an angular range of $\sim 8^{\circ}-26^{\circ}$ in laboratory system.
	Finally, tritons originating from the $^{7}$H decay and moving in a narrow cone of forward angles, $\theta_t \leq 6^{\circ}$, were detected by the $61\times 61$ mm$^2$ telescope which was installed at zero laboratory angle $280$ mm downstream from the target.
	It consisted of one 1500-micron thick Si DSD ($32\times 32$ strips) and a set of 16 square CsI(Tl)/PMT modules (the CsI(Tl) crystals were 50 mm thick).
	The $^{3}$H telescope provided angular resolution of $\sim 0.5^{\circ}$ and energy resolution of $\sim 2\%$.

	\begin{figure}[t]
	\begin{center}
	\includegraphics[width=0.45\textwidth]{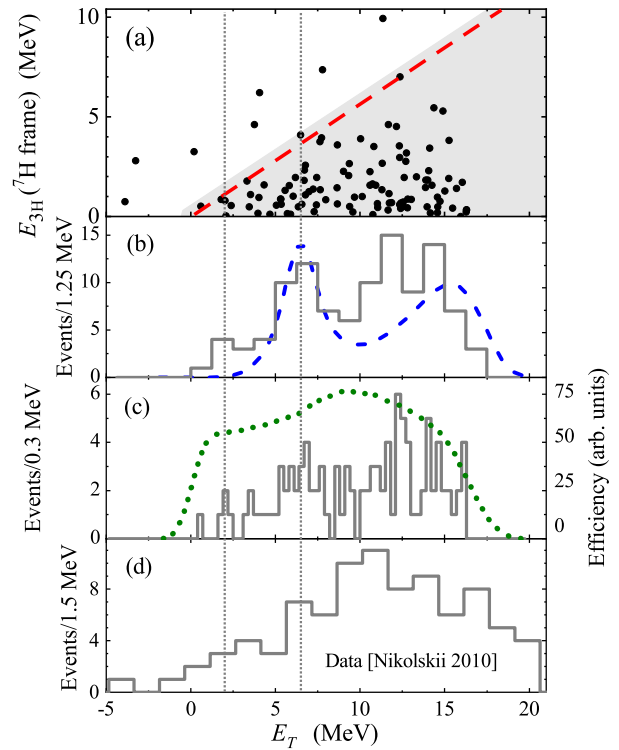}
	\end{center}
	\caption{Missing mass spectrum of $^{7}$H obtained for the $^2$H($^8$He,$^3$He)$^7$H reaction with the $^3$He recoils detected in coincidence with tritons.
	Panel (a) shows correlation between the $^3$H energy in the $^{7}$H c.m. frame and the $^{7}$H MM energy.
	Red dashed line shows the kinematical limit for tritons coming from $^{7}$H decay.
	Grey triangle represents kinematicaly allowed region used for the reconstruction of the MM spectrum.
	Panels (b) and (c) show the MM spectrum with two different binning factors.
	Blue dashed curve in panel (b) shows the simulation of the 6.5 MeV state with $\Gamma=2$ MeV plus the contribution of the $t$+$4n$ five-body phase volume (arbitrary normalization) convoluted with the experimental setup efficiency and resolution.
	Green dotted curve in panel (c) shows the experimental setup efficiency for the $^{7}$H registration.
	Panel (d) shows the data from Ref.\ \cite{Nikolskii:2010}.
	Vertical dotted lines indicate the presumed positions of the $^{7}$H ground and first excited states.
	}
	\label{fig:mm}
	\end{figure}

\textit{Missing mass spectrum.} --- %
	%
	All together 107 events were detected in the experiment. Fig.\ \ref{fig:mm} (a) shows correlation plot between the $^{7}$H MM and $^3$H energy in the $^{7}$H center-of-mass (c.m.)\ frame.
	It can be seen that the majority of data is in agreement with the hypothesis of $^{7}$H population and its subsequent decay.
	The events outside the kinematicaly allowed region are very few and evenly distributed.
	The MM spectrum of $^{7}$H is shown in Figs.\ \ref{fig:mm} (b), (c) in different representations.
	In this spectrum the peak with energy $E_T=6.5(5)$  MeV, width $\Gamma=2.0(5)$ MeV, and population cross section of $\sim 30$ $\mu$b/sr can be well identified.
	This peak is interpreted as the first excited state of $^{7}$H, though the $5/2^+$ and $3/2^+$ doublet of the lowest excited states cannot be excluded.
	There is also a compact group of events at $E_T \sim 2$ MeV emerging at $^{7}$H c.m.\ angles $17^{\circ}-27^{\circ}$.
	This group has population c.m. cross section of $10$ $\mu$b/sr, and we associate it with the $^{7}$H ground state.
	Such an interpretation is at the limit of statistical significance and deserves special discussion.
	Figs.\ \ref{fig:mm} (b), (c) show that the MM spectrum at $E_T>12$ MeV can be explained by the combination of rapidly growing 5-body phase volume and rapidly falling detection efficiency.
	
	\begin{figure}[t]
	\begin{center}
	\includegraphics[width=0.43\textwidth]{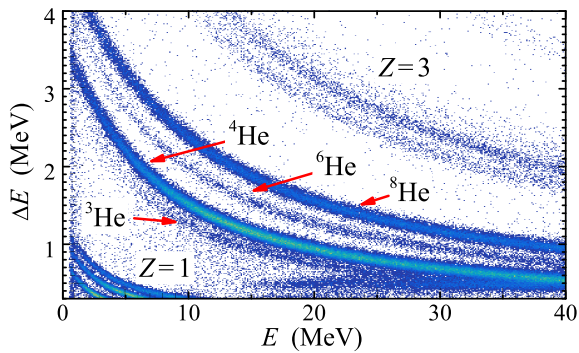}
	\end{center}
	\caption{The $^{3}$He recoil identification. 
		The accurate thickness mapping of the $\Delta E$ 20 $\mu$m detectors was made before the beam exposition 
		to get reliable $^{3}$He-$^{4}$He separation.}
	\label{fig:hyperb}
	\end{figure}

\textit{Discussion of the $^{7}$H ground state evidence.} --- %
%
	We consider the group of events with $0.5<E_T<2.5$ MeV as candidate for the $^{7}$H ground state.
	Because of small statistics (5 events) this group can be regarded only as an indication of the possible ground state.
	To increase confidence in this interpretation, let us consider all the candidate events in details.

	Fig.\ \ref{fig:hyperb} demonstrates the good quality of the $^{3}$He recoil identification.
	It is clear that the $^{3}$H fragment identification in the zero-angle telescope is much better.
	Thus the decay channel identification is unambiguous for all the events in Fig.\ \ref{fig:mm} (a).
	The channel identification was checked especially carefully for the individual g.s.\ candidate events.
	It can be also seen in Fig.\ \ref{fig:mm} (a) that all the $^{7}$H g.s.\ candidate events are located within the kinematical locus associated with the $^{7}$H decay hypothesis.

	\begin{figure}[t]
	\begin{center}
	\includegraphics[width=0.23\textwidth]{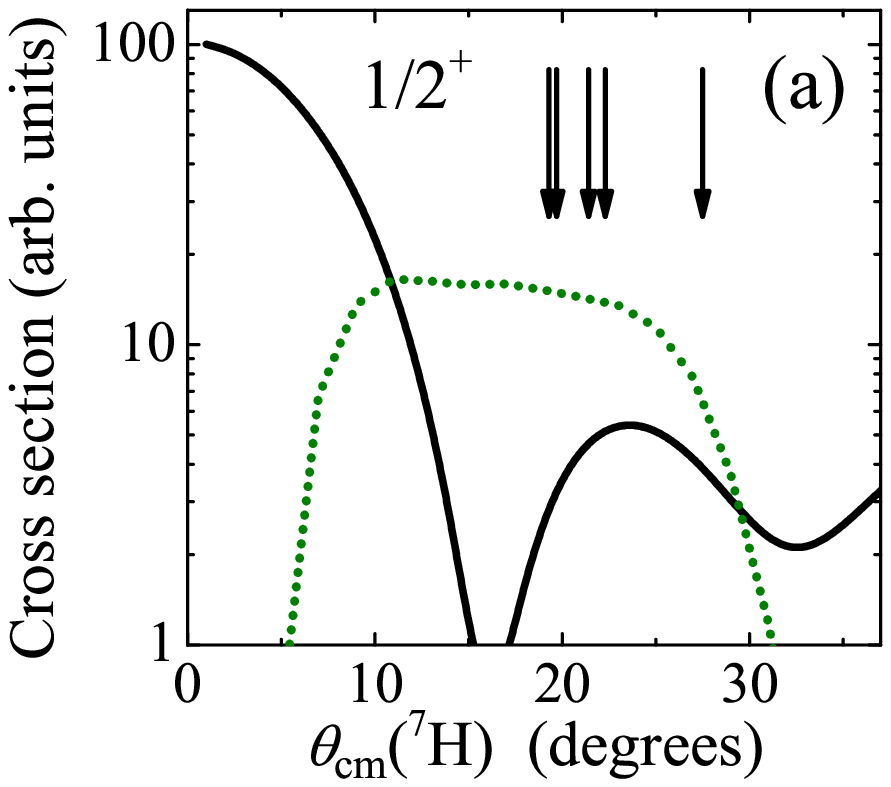}
	\includegraphics[width=0.238\textwidth]{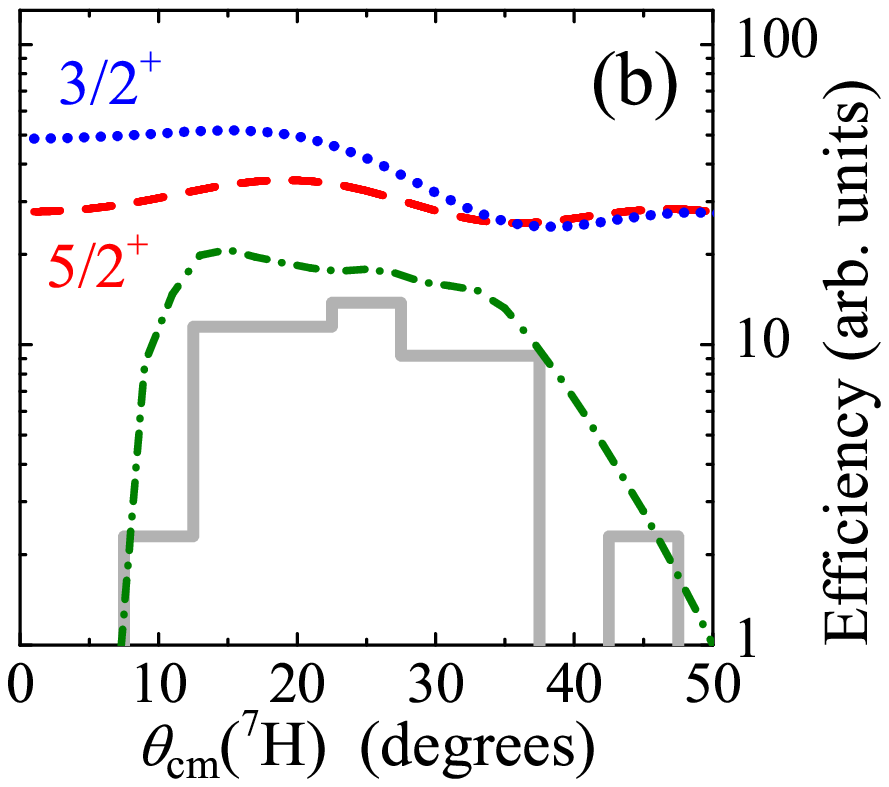}
	\end{center}
	\caption{The center-of-mass angular distributions calculated with the use of the FRESCO code for the $^2$H($^8$He,$^3$He)$^7$H reaction channels leading to the formation of $^{7}$H in the three states.
	Calculation results obtained for the $1/2^+$, $5/2^+$, and $3/2^+$ states are shown with the black solid (a), red dashed (b) and blue dotted (b) curves, respectively.
	The angular distributions presented for the $5/2^+$ and $3/2^+$ states were obtained assuming the entrance-channel two-step transition through the $^{8}$He $2^+$ excitation.
	The angles of the five events associated with the population of the $^7$H ground-state candidate are indicated by arrows in (a).
	The gray histogram in (b) shows the angular distribution of events in the $5-8$ MeV energy range around the $E_T=6.5$ MeV the excited state position.
	The green dotted and dash-dotted curves, associated with the right axis, show the experimental setup efficiency for the registration of produced $^{7}$H with $E_T$ equal to 2 and to 6.5 MeV, respectively.}
	\label{fig:cm-angdis}
	\end{figure}
	
	The angles of the respective $^{7}$H g.s. candidate events are shown by arrows in Fig.\ \ref{fig:cm-angdis} (a).
	Our setup was not suited for the $^{7}$H g.s.\ detection in the forward-angle cross-section maximum of the $^2$H($^8$He,$^3$He)$^7$H reaction.
	It can be seen that all the g.s.\ candidate events are concentrated in the region predicted to be the second diffraction maximum for the calculated cross section of the $1/2^+$ state.
	
\begin{figure}[t]
	\begin{center}
	\includegraphics[width=0.232\textwidth]{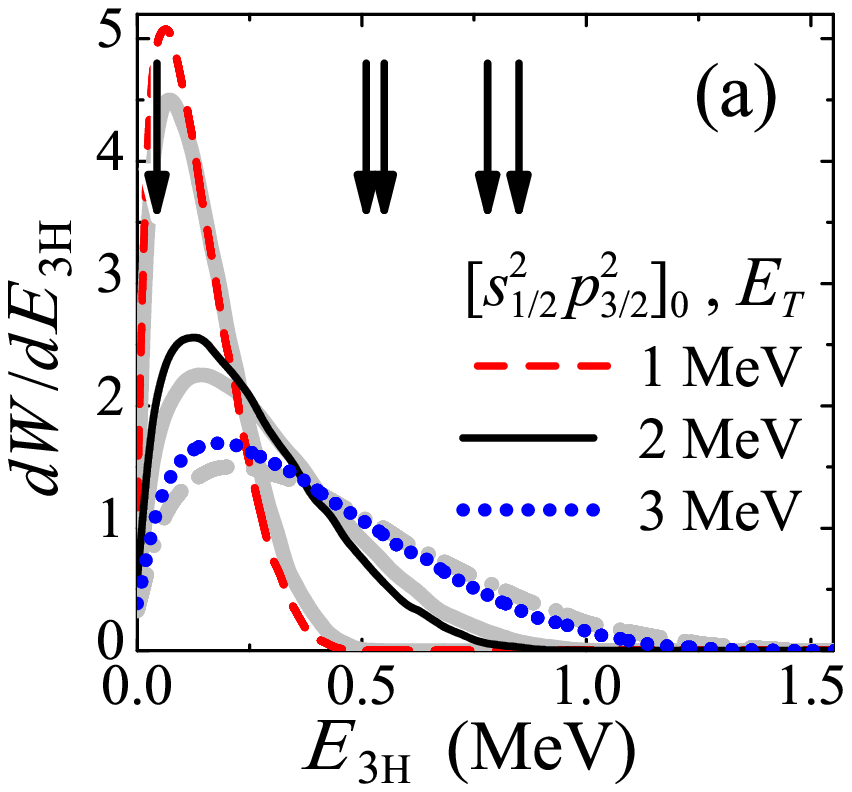}
	\includegraphics[width=0.238\textwidth]{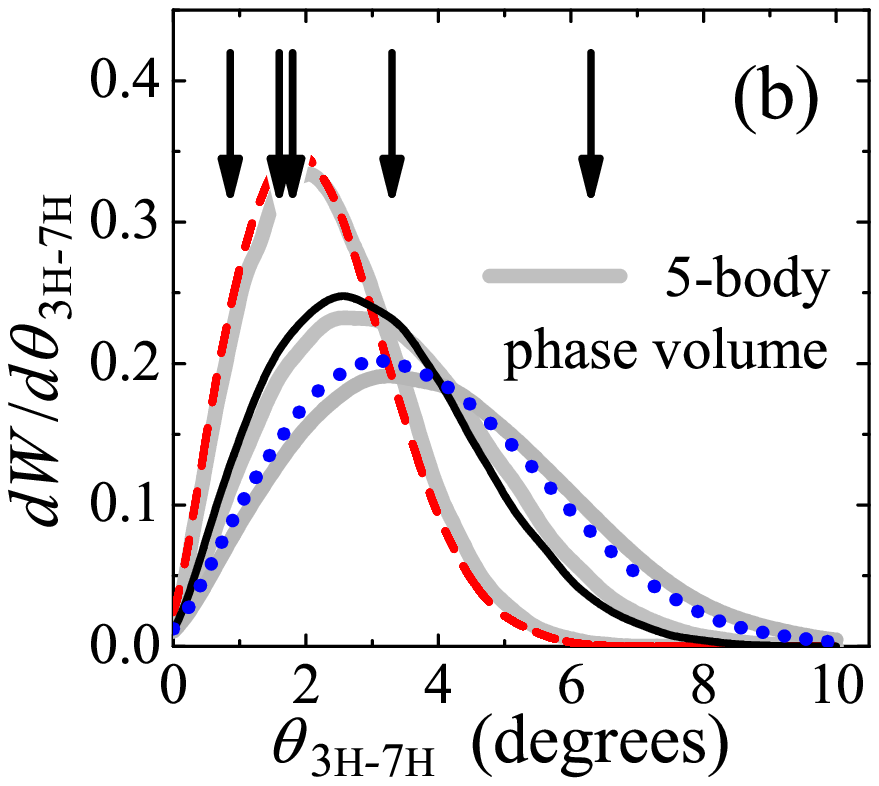}
	\end{center}
	\caption{The solid, dashed and dotted curves correspond to the decay simulations performed for the $[s^2_{1/2}p^2_{3/2}]_0$ valence neutron configuration of $^{7}$H at different decay energies $E_T$.
	This configuration has the best barrier penetration properties, and it is expected to dominate in the decay of $[p^4_{3/2}]_0$ internal structure, see \cite{Sharov:2019} for details.
	The thick gray curves show the corresponding 5-body phase-space distributions.
	(a) Energy distribution of $^{3}$H produced in $^{7}$H decay in the $^{7}$H frame.
	(b) Angular distribution of $^{3}$H relative to the reconstructed $^{7}$H flight 	direction in the laboratory frame.
	The angles of the five candidate events associated with the population of the $^{7}$H g.s.\ are indicated by arrows.}
	\label{fig:t-distr}
	\end{figure}

	Correlation patterns anticipated for the decay of the core+$4n$ systems were studied in the recent paper \cite{Sharov:2019}.
	Among the correlations considered in \cite{Sharov:2019} only the energy distribution of $^{3}$H in the $^{7}$H frame can be reconstructed from experimental data of the present work.
	These distributions are expected to have profile with quite a narrow low-energy peak.
	Their shape can be affected by the decay dynamics of $^{7}$H as well, see Fig.\ \ref{fig:t-distr} (a).
	For comparison with the measured data we calculated also the angular distribution of $^{3}$H relative to the reconstructed $^{7}$H flight direction in laboratory frame, see Fig.\ \ref{fig:t-distr} (b).
	In contrast with the $^{3}$H energy in $^{7}$H frame, the mentioned angle is defined with much higher precision.
	Namely, the $^{3}$H direction is defined with precision of $ 0.5^{\circ}$ by the forward telescope, and the $^{7}$H direction is deduced with precision of $ 0.5^{\circ}$ based on the momentum vectors of the incoming $^{8}$He beam and the $^{3}$He recoil.
	It can be seen in Fig.\ \ref{fig:t-distr} (b) that all the candidate g.s.\ events nicely fit in the theoretically predicted angular distribution peak.
	
	The $^{7}$H g.s.\ position at $E_T=2.0(5)$ MeV, suggested here, is consistent with the observation of near-threshold anomaly in Ref.\ \cite{Korsheninnikov:2003}.
	Our spectrum of $^7$H for $E_T < 8$ MeV is consistent with the spectrum of Ref. \cite{Nikolskii:2010}, see Fig. \ref{fig:mm} (d).
	The latter was obtained in the same reaction at different energy and with worse energy resolution.
	The g.s. energy inferred in our work strongly differs from the value  $E_T=0.57^{+0.42}_{-0.21}$ MeV 	reported in \cite{Caamano:2007}, far beyond the declared experimental errors. 
	Another subject of concern is the the large cross section reported in \cite{Caamano:2007} for the $^{12}$C($^{8}$He,$^{7}$H)$^{13}$N reaction populating the $^7$H g.s., while this reaction is less preferable than the ($d$,$^{3}$He) reaction, e.g.\ due to the $Q$ value (see also discussion of this issue in \cite{Nikolskii:2010}).

\textit{Discussion of the $^{7}$H excited state.} --- %
%
	What can be the nature of the 6.5 MeV state in $^{7}$H? It should be noted that $^{7}$H has closed $p_{3/2}$ subshell.
	Systems with shell closure typically have quite poor low-lying excitation spectrum, and the easiest expectation is that the lowest is the $2^+$ excitation formed by pushing neutrons to the $[p^2_{3/2}p^2_{1/2}]_2$ configuration.
	The $2^+$ excitation of valence neutrons should be coupled with core spin to the $5/2^+-3/2^+$ doublet.
	The separation of the doublet members is questionable, and here we can refer only to the experience of the $^{5}$H excited states' studies in Ref.\ \cite{Golovkov:2005} where this separation was found to be insignificant.

	\begin{figure}[t]
	\begin{center}
	\includegraphics[width=0.47\textwidth]{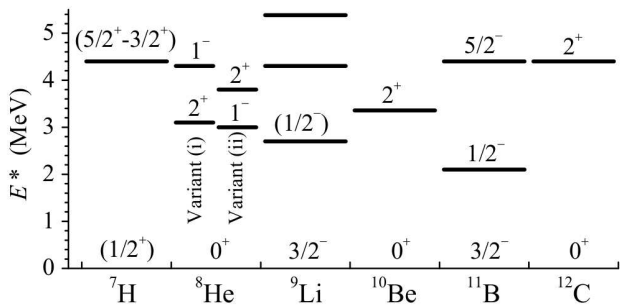}
	\end{center}
	\caption{Systematics of the lowest excited states for the isotones with closed $p_{3/2}$ neutron subshell.
	For $^{8}$He there exist two different prescriptions of the low-lying spectrum: (i) $2^+$, $1^-$ \cite{Markenroth:2001,Meister:2002} and  (ii) $2^+$, $1^-$ \cite{Golovkov:2009}.}
	\label{fig:en-syst}
	\end{figure}
	
	The systematics of the lowest excited states for light systems with closed $p_{3/2}$ is given in Fig.\ \ref{fig:en-syst}.
	It can be seen that excited states which can be related to the excitations of the neutron configurations have typical energies $E^* \sim 3.5-4.5$ MeV.
	In this plot the $^{7}$H excitation energy is determined assuming that the group of events at $E_T=2.0$ MeV represents the g.s.\ position, which gives excited state position $E^* \sim 4.5$ MeV, fitting well the systematics.
	If we admit lower $E_T$ values for the g.s., for example $E_T< 1$ MeV, we get unexpectedly high energies for the $^{7}$H excited state, $E^* > 5.5$ MeV.
	This can be considered as additional argument supporting our prescription of the $^{7}$H g.s.
	
	The $^{7}$H c.m.\ angular distribution for the 6.5 MeV excitation region is shown in Fig.\ \ref{fig:cm-angdis} (b).
	The experimental angular distribution corresponds well to the $5/2^+$ and $3/2^+$ distributions calculated by FRESCO code \cite{FRESCO} with the setup efficiency taken into account.

\textit{Conclusion.} --- %
%
	The following major results are obtained in this work:
	
	\noindent (i) For the first time, the $^7$H excited state is observed at $E_T = 6.5(5)$ MeV with $\Gamma = 2.0(5)$ MeV.
	This state can be interpreted as unresolved $5/2^+$ and $3/2^+$ doublet, built upon the $2^+$ excitation of valence neutrons, or one of the doublet states.
	
	\noindent (ii) Indications for the $^7$H g.s.\ at $E_T = 2.0(5)$ MeV are found in the measured energy and angular distributions.
	
	\noindent (iii) The measured c.m. population cross section of the presumed $^7$H g.s.\ is about 10 $\mu$b/sr, which clarifies why the previous searches for the $^7$H g.s.\ required so much time and effort without bringing reliable assignments of such a remote isotope.
	
	The obtained results represent an important step towards resolving the problem of the $^{7}$H observation and also demonstrate the high potential of the ``newcomer'' ACCULINNA-2 facility.

\textit{Acknowledgments.} --- %
%
	We acknowledge important contribution of Prof. M.S. Golovkov to the development of the experimental setup.
	This work was supported in part by the Russian Science Foundation grant No.\ 17-12-01367 and MEYS Project (Czech Republic) LTT17003.


\bibliography{all.bib}


\end{document}